# CHESS Compact Wiggler construction report.

Alexander Temnykh[1*] and Ivan Temnykh [2]
[1]CHESS, Cornell University, Ithaca, NY 14850, USA;
[2] Pine Hollow Auto Diagnostics, Pennsylvania Furnace, PA 16865, USA
* Corresponding author, E-mail: abt6@cornell.edu

## Abstract

We developed, built, characterized on bench and beam-tested a permanent magnet (PM) Compact Wiggler (CW) prototype with a hydraulic assist gap-controlling mechanism. The prototype is of ~50cm long, 20cm wide and 40cm high and weights ~50kg. Magnetic structure has a 76.2mm period. At 6.5mm minimal gap the structure demonstrated ~2.3 Tesla peak field. At this gap, the magnet arrays attract each other with 7,990 N (1,796 lbf) force. Hydraulic system is employed to balance the attractive forces.

Here we describe the mechanical and magnetic design, present some construction details and provide test results.

## 1. Introduction

The development of CW was motivated by intention to improve performance of CHESS 1A2 & 1A3 beamlines used for Structural Materials studies.

Presently, 1A2 & 1A3 beamlines utilize radiation generated by a very old 25-pole hybrid permanent magnet wiggler built in 1990 [1]. At that time, CESR operated as an electron-positron collider. Operation required quite large vertical aperture as well as big "good-field-region" and a large aperture in the horizontal plane. That obligated the wiggler to have 40mm minimal gap and 100 mm pole width. After CHESS-U upgrade [2] in 2018 and switching to single-beam operation, requirements for insertion devices changed. Presently, minimal gap can be as small as 6.5mm and pole width around 25mm. New requirements as well as new ideas (see [3,4,5]) enabled us to design and construct a Compact Wiggler (CW) of much better performance in terms of photon flux at high energy.

The following sections describe mechanical and magnetic designs, construction details and bench and beam testing results of the ~50cm long Compact Wiggler prototype.

## 2. CW prototype mechanical design.

Mechanical design of CW prototype is similar to the design of sCCU-type undulators (see [3] and [5]).

It consists of a stationary frame (1), two movable girders with magnet arrays (2) coupled to the frame with sliders (3) and mechanical drivers (4) (Fig. 1). The hydraulic system, composed of 16 miniature cylinders (5) connected in series, and a hydraulic pump, helps the mechanical drivers to move the girders and provides magnetic force compensation. The pressure in the hydraulic system is regulated by the automatic control system and depends on the wiggler gap (i.e. on magnetic forces).

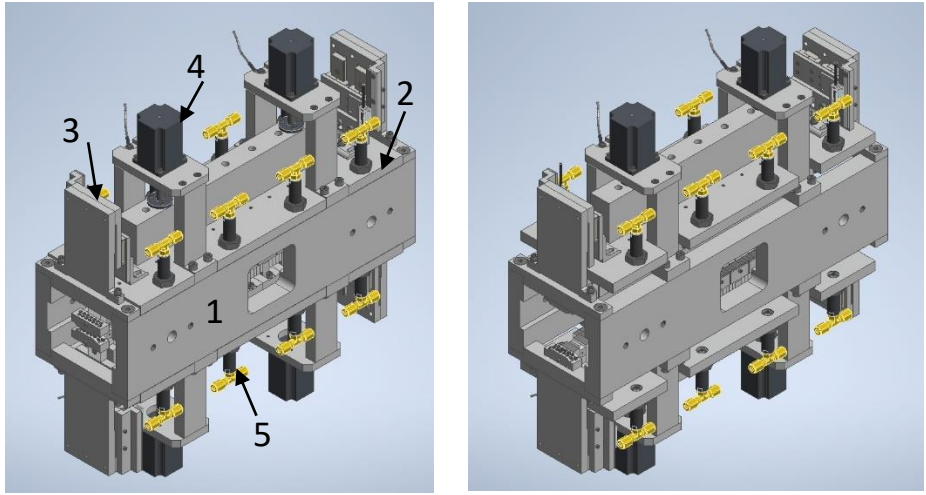

*Figure 1. Compact Wiggler prototype. On the left – "Inventor" model with gap is closed; on the Right – model with gap open. Here are: 1 – stationary frame; 2- movable girders; 3- sliders coupling girders with frame; 4- mechanical drivers; 5- miniatuire hydraulic cyliders, connectiong tubing is not shown.*

Because the CW magnetic field is ~3 times stronger than it is for sCCU28 [2], the hydraulic system was slightly modified as described below.

At minimal 6.5mm gap, calculations predicted 7,990 N (1,796 lbf) attracting forces between magnet arrays. Because of hydraulic component dimensions and system layout, the distance between hydraulic cylinders cannot be less than ~130mm. This limits the maximum number of cylinders which can be fitted into the girder. For ~540mm-long girders, it is possible to fit 4 cylinders on each side or 8 total (Fig. 1). To balance 7,990 N (1,796 lbf) of magnetic forces, each cylinder should provide 7,900 N /8 ~998 N (224 lbf) pushing force.

In sCCU28 and sCCU19 projects we used Vektek model 20-0104-07 hydraulic cylinders with ~0.11 sq. in. piston area. If we used the same cylinders for CW prototype, the magnetic forces balance at minimal gap would require 2,036 Psi hydraulic pressure. To have pressure in a range similar to what we had in previous projects, we opted to employ cylinders with a larger 0.196 sq.in piston area (Vektek model 20-0105-07). That resulted in a moderate maximum pressure of ~1,100 psi. From previous projects we learned that at this pressure, all the hydraulic components work very reliably.

The measured and predicted pressures in the hydraulic system required for balance of magnetic forces between arrays as a function of wiggler gap are in good agreement (Fig.2).

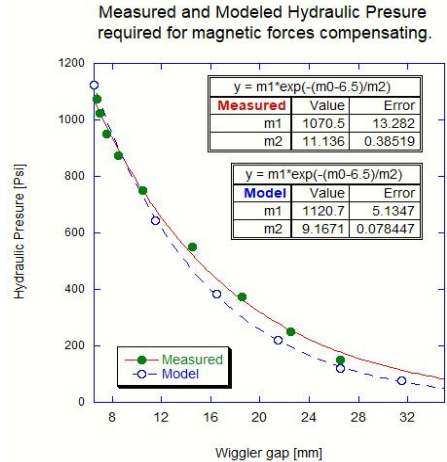

*Figure 2. Measured and calculated hydraulic pressure required for magnetic force balance.*

Note, all other components such as sliders, mechanical drivers, pressure transducer, load cells, electronics modules etc., are identical to what we used in sCCU28 and sCCU19 projects.

## 3. CW prototype magnetic structure.

The CW prototype magnetic structure consists of two, top and bottom, magnet arrays. Each array, comprises 14 single pole assemblies with alternating magnetic field (Fig. 3). A single pole assembly consists of an aluminum base, a magnetic field concentrator made of VacoFlux alloy (know also as Hiperco 50 and Permendur 2V) with high saturation magnetic field and 13 NbFeB permanet magnet blocks arranged around concentrator (Fig.3 Left, Middle). Red arrows show direction of PM blocks polarization. For indicated polarization, magnetic field on top of the concentrator will be directed *up*. Poles with magnetic field directed *down*, have PM blocks polarization in opposite direction.

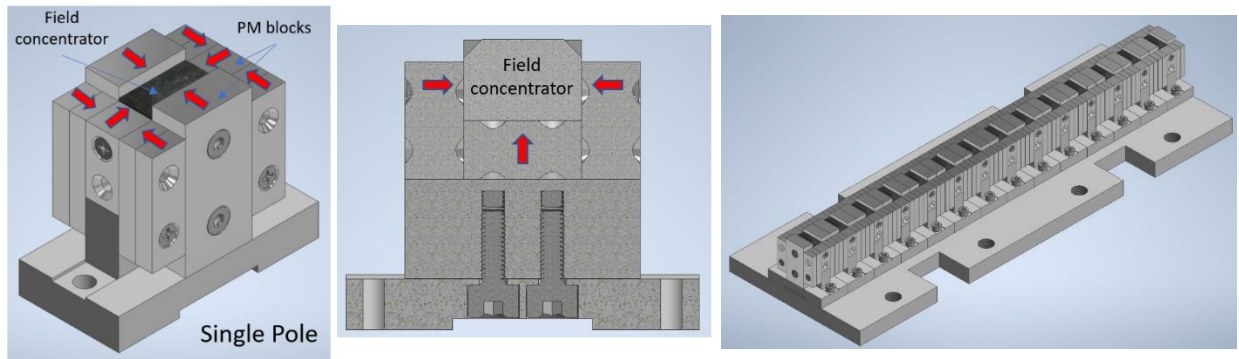

*Figure 3. On the left - single pole unit; in the middle – single pole cross-section; on the right – CW magnet array assembly.*

Three types of PM blocks used in assembly were purchased from "K&J Magnetic, Inc." (https://www.kjmagnetics.com/) (Fig. 4). Part number and dimensions are given in the table below.

| PM block dimension | Number of blocks | Part Number |
| --- | --- | --- |

| W x H x T | per pole | |
|---|---|---|
| 2" x 1" x ½ " | 2 | BY0X08DCS-N52 |
| 1" x ½" x ½" | 3 | BX088DCS |
| 1 ½" x ½" x ¼" | 8 | BX884DCS |

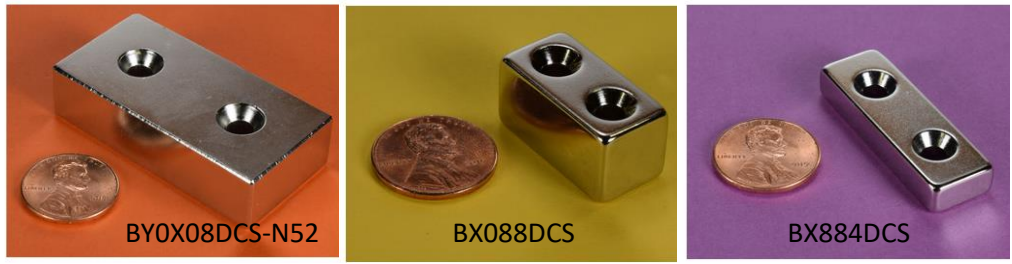

*Figure 4. Three types of PM blocks used in CW construction.*

For fastening were used #6 and #8 flat head screws from McMaster (https://www.mcmaster.com/)

Field concentrators were made from VacoFlux material left over from PM wiggler construction dating back to 1991.

To assemble the individual poles (Fig. 5), first the field concentrator (1) and one PM block (2) was attached to the base. The colors on the concentrators indicate magnetic field direction (red-up and blue-down). The fully assembled pole unit contains all PM blocks fastened to the base and field concentrator by stainless steel screws.

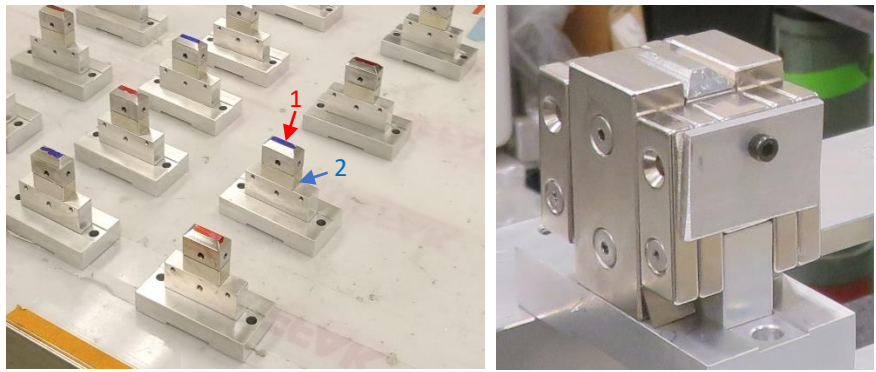

*Figure 5. Pole assembly illustration. On the left - BX088DCS PM blocks (2) and field concentrators (1) attached to base block. On the right – fully assembled unit.*

3D magnetic field modeling was done with OPERA Simulation Software (Fig. 6). The model predicted 2.3T peak field at 6.5mm gap; the field roll-off data was also plotted. The fitting coefficients can be used for dynamic field integrals calculation.

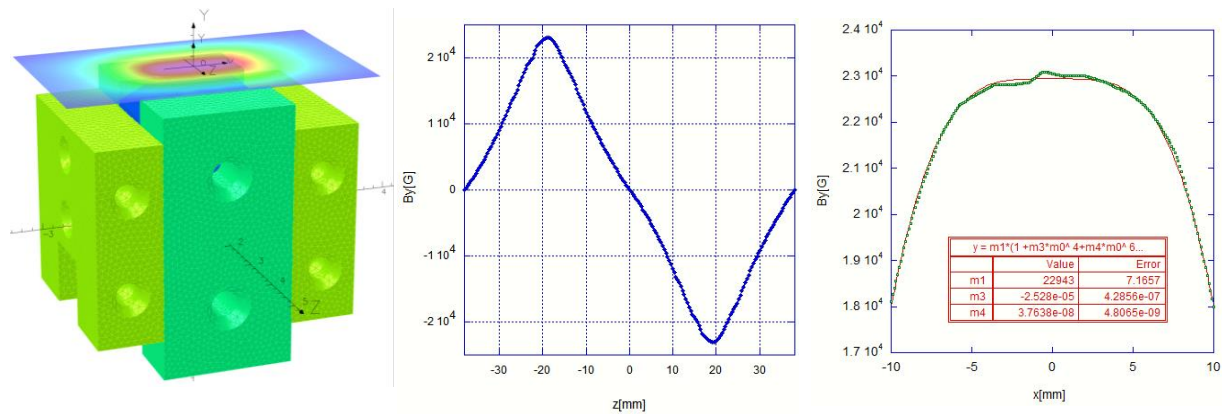

*Figure 6.3D magnetic field simulation results. On the left – model used in simulation; in the middle – one period of vertical magnetic field along beam axis, simulation predicted ~2.30T peak field; on the right – field roll off across the pole data, the fitting coefficients can be used for dynamic integrals calculation.*

## 4. Inventor model and drawing location.

CW prototype Inventor model as well as production drawing can be found in CHESS internal file system in Vault in folder "…6000\6076\6076-200 Compact Wiggler\6076-202 Compact Wiggler MKII"; 3D model in file "CCW_Mark2_FullModel_SWH.iam" and complete set of drawings in "6076-202.idw". The Vault is

## 5. CW magnetic field tuning and final characterization results.

Magnetic field tuning and characterization was done on the same bench used for undulators field measurements (see [2]). Procedure and characterization results are presented and discussed below.

### 5.1. Magnet arrays tuning.

We measured the magnetic field properties of the individual arrays before final assembly using a procedure similar to that used for undulator construction. We installed and aligned the arrays, one at a time, on the magnetic field measurement bench. We used the scanning Hall probe above poles at a height corresponding to the beam axis to measure the magnetic field profile along beam axis and map the field in the horizontal plane (*Upper* - Fig. 8 and *Lower* - Fig. 9).

The *Upper* array magnetic field profile along the beam axis is asymmetric with respect to center and has 14 peaks, with each peak corresponding to a single pole (Fig. 8 Left). The beam trajectory was calculated using the measured field profile (Fig. 8 Middle).

Note: the insertion device (ID) magnetic structure should provide a smooth transition between trajectories inside and outside of ID. Usually, that is done by modification of two poles at each ID end. In our case, we reduced the strength of four poles, two at entrance and two at exit, at locations indicated by red arrows (Fig. 8 Left). The pole strength was reduced by removing the appropriate number of PM blocks. This adjustment resulted in a smooth trajectory transition (Fig. 8 Middle). A satisfactory transition was confirmed by the beam test described in Section 6. 2D field mapping confirmed off-axis field had no issues (Fig. 8 Right).

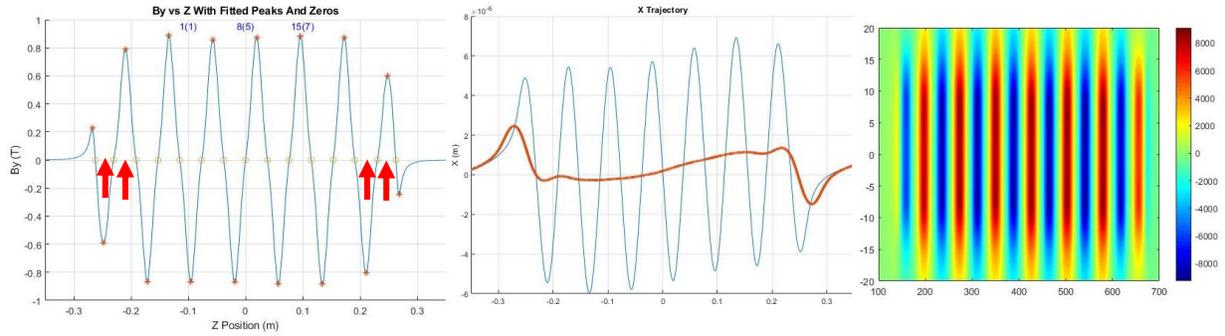

*Figure 8. On the left - Top CW array on bench; in the middle - vertical field profile, red arrows indicate location of modified poles; on the right - beam trajectory corresponding measured field profile, red plot is trajectory averaged over one structure period.*

The *bottom* array was assembled and tuned in the same fashion. We measured the magnetic field profile was measured along beam axis, plotted the predicted trajectory with smooth transition and performed 2D field mapping to confirm off-axis field properties (Fig. 9).

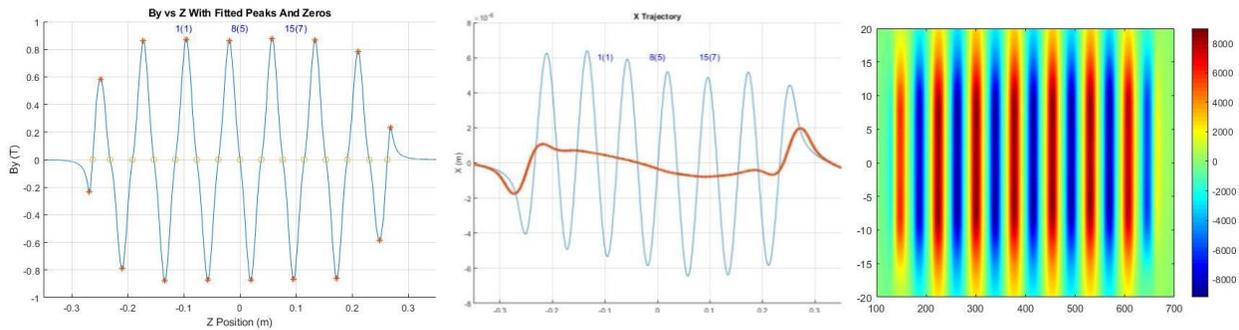

*Figure 7. On the left - bottom array field profile along beam axis; on the right – beam trajectory corresponding measured field.*

### 5.2. Assembled device magnetic field characterization.

We measured the magnetic field of the assembled device using the standard method (see [3]). Two rails (1) attached to side plates of the stationary frame guide the carriage (2) with Hall sensor along the beam axis (Fig. 10). The carriage was moved along prototype with carbon fiber rods while vertical magnetic field was measured with Hall sensor and recorded. Prior to the measurements, the Hall sensor was accurately calibrated against NMR probe with ~2e-4 precision.

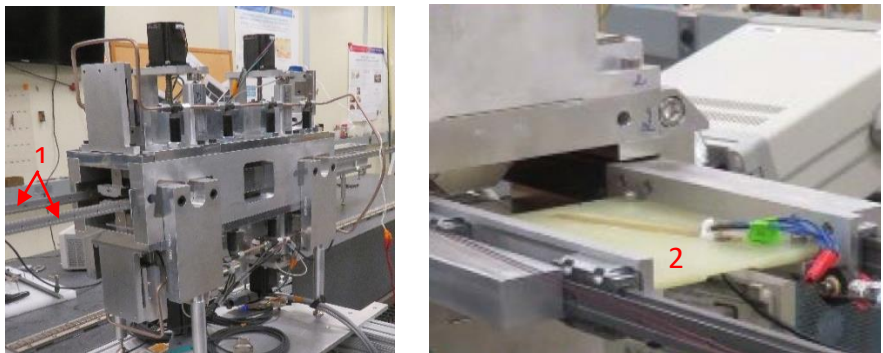

*Figure 9, Assembled CW magnetic field measurement setup. On the left – assembled CW prototype on the bench. Here "1" indicates rails attached to CW frame and used to guide Hall probe carriage. On the right – Hall probe carriage with Hall sensor.*

We plotted the measured field at minimal (closed) 6.5mm gap and calculated beam trajectory (Fig. 11). At this gap, measurements indicated 2.29 T peak field: very close to 2.30 T peak field predicted by model. Beam trajectory calculated using the measured field revealed negligible deflection and ~15 microns horizontal orbit offset (Fig. 11 Right). The 15micron offset in horizontal plane was considered acceptable: it is much smaller than horizontal beam size and it will be easily corrected by automatic orbit correction process.

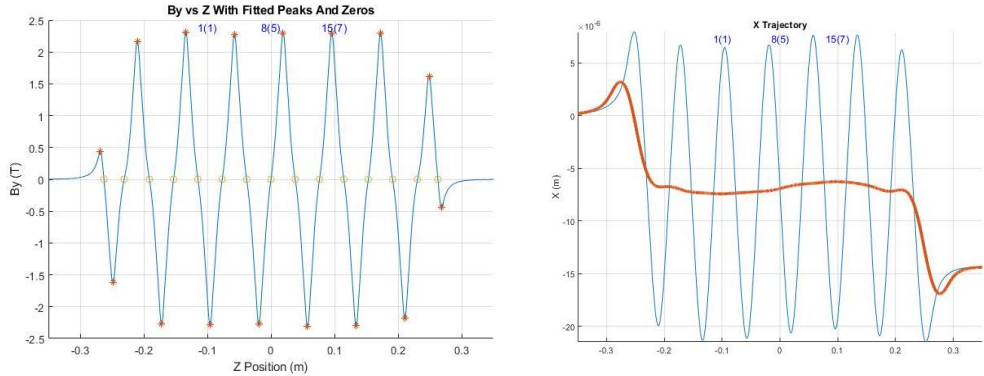

*Figure 10. Magnetic field profile along CW (on the left) and calculated beam trajectory (on the right) at minimal (closed) 6.5mm gap.*

Although in operation the CW gap will be closed and field will be kept constant, for better understanding of magnetic structure properties we measured field strength as a function of gap (Fig. 12). Here the average peak field is plotted as a function of gap. The field dependence on gap was fitted with exponential function:

$$B_{peak}[T] = m_1 \times \exp\left(-\frac{(Gap - 6.5)}{m_2} + \frac{(Gap - 6.5)^2}{m_3}\right)$$

where $B_{peak}$ is in Tesla, $Gap$ – in mm. The fitting yielded: $m_1 = 2.289$; $m_2 = 12.009$ and $m_3 = 807$.

### 5.3. Field integrals correction.

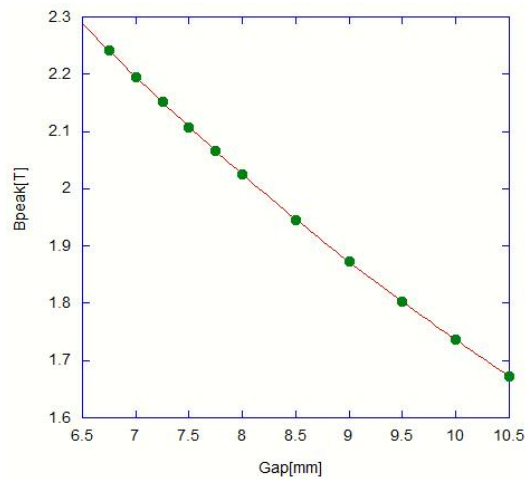

*Figure 11. CW measured average peak field as a function of gap.*

Variation or dependence of ID field integrals on offset from beam axis negatively affects the beam dynamic in the storage ring. To avoid this, the variation of integrals should be minimized. Dependence of vertical and horizontal field integrals (Ix, Iy) of the Compact Wiggler prototype on horizontal position was measured with long coils stretched through the undulator and corrected with the "magic fingers" technique (see [2]). Initial measurement of the field integrals indicated their variation of approximately $\pm 8.1$ G-m range. After few iterations of measurements and MF adjustments (Fig. 13), horizontal and vertical field integrals variation was reduced to acceptable $\pm 1.8$ and $\pm 1.1$ G-m levels.

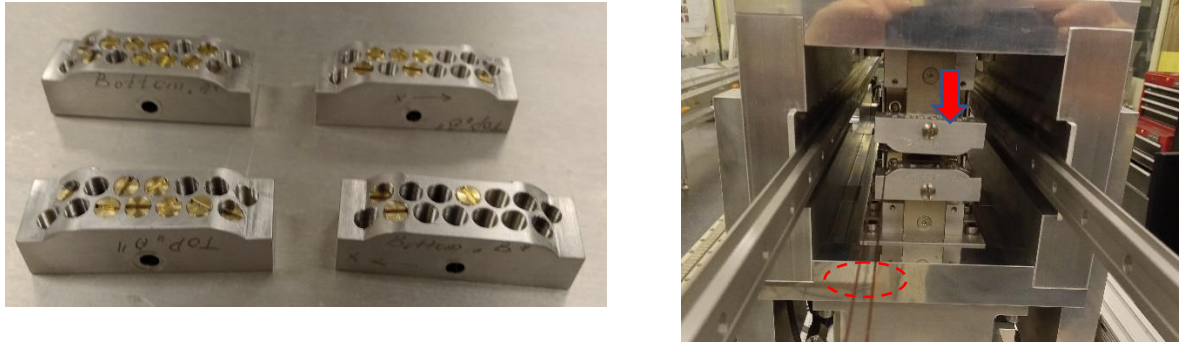

*Figure 12. On the left - magic fingers blocks used for field integrals correction; on the right – CW end view, red arrow points on MF block attached to magnet array end and oval marks "long coil" used for field integrals measurement.*

## 6. CW prototype beam test results

The main goal for the beam test was to verify that CW would not create any problems to CESR/CHESS operation.

It should be noted, narrow wiggler poles promoting fast field roll-off in combination with strong peak field give rise to dynamic field integrals. That, in turn, has created concerns about dynamic aperture limitation and potential problems with injection efficiency.

To resolve these concerns, CW prototype was installed in Sector 4 in April 2025 (see CHESS internal, CHESS Elog 04/08/2025) by CHESS engineers Timothy Joyce and Keith Surrena. Shortly after installation, Accelerator Physicist Suntao Wang conducted the beam test (see CHESS internal, CHESS_MS 4/11/2025).

Below is the test summary copied from the MS report:

---

a) With CW (Compact Wiggler) closed, single bunch injection is not affected (~50%) after orbit correction and injection tuning
b) Initial multiple bunch injection is comparable with the condition with CCW open at low current. At 145mA, the injection efficiency goes down from 33% to 24%.
c) Tune shift is small <0.3kHz. Beam life time, radiation is comparable with CCW open.
d) This 0.5m prototype CCW has minimum effect on CESR operation. CESR can run with this CCW.

---

These results validated CW design and confirmed that this device is well-suited for operation at CHESS/CESR.

## Conclusion.

0.5m long Compact Wiggler prototype with 2.3T magnetic peak field and hydraulic assisted gap controlling mechanism was designed, constructed, characterized on bench and successfully tested with beam in storge ring. Results of bench characterization confirmed design parameters; successful beam test validated the design concept.

The use of a full-length (1.5m) Compact Wiggler as a SR source for ID1A beam line would increase photon flux by a factor from 5 to 30 depending on photon energy as illustrated in Appendix. That would significantly boost the beam line performance,

## Acknowledgement

The authors would like to thank Research Scientist Arthur Woll for motivation and support of CW construction. Special thanks to CHESS Engineers Timothy Joyce and Keith Surrena for CW preparation for beam test and to Accelerator Physicist Suntao Wang for conducting the beam test and analyzing the test results.

## References

[1] K. D. Finkelstein, The new CHESS wiggler, Rev. Sci. Instrum. 63, 305–308 (1992), https://doi.org/10.1063/1.1142976

[2] CHESS-U Upgrade

[3] Alexander Temnykh and Ivan Temnykh, sCCU—Compact Variable-Gap Undulator with hydraulic-assist driver and enhanced magnetic field, NIMA 1039 (2022) 167091 https://doi.org/10.1016/j.nima.2022.167091

[4] US Patent Number: US-12119130-B2

[5] Alexander Temnykh and Ivan Temnykh, Short-period compact undulator (sCCU19) construction report, https://doi.org/10.48550/arXiv.2501.02391

## Appendix.

Example of photon flux simulation for "old" F-line and "new" 1.5m Compact wigglers. Phonon flux was calculated for 1mm x 1mm aperture 50m away from source and 145mA beam. Plot "a" shows flux with no filters or absorbers; plot "b" is the simulation results for absorber included 0.38 mm of C, 0.686 mm of Be, 3.48 mm of Al, 0.487 mm of Si and 1 mm of Cu. In both cases, one can expect significant gain in photon flux by using the Compact Wiggler.

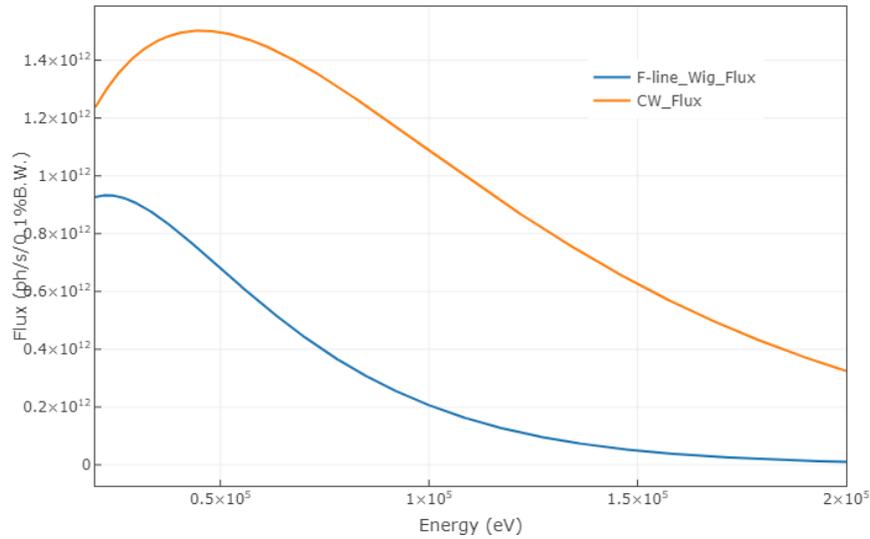

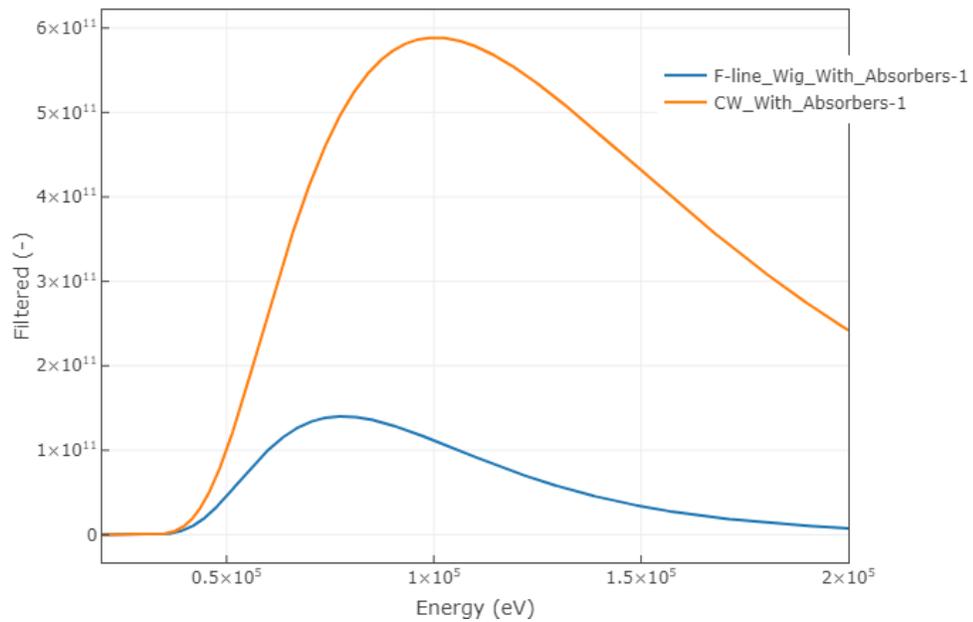